# Hybrid-order topological insulators in a phononic crystal

Yating Yang[1*], Jiuyang Lu[1*], Mou Yan[1], Xueqin Huang[1†], Weiyin Deng[1†], and Zhengyou Liu[2,3†]

[1]School of Physics and Optoelectronics, South China University of Technology, Guangzhou 510640, China

[2]Key Laboratory of Artificial Micro- and Nanostructures of Ministry of Education and School of Physics and Technology, Wuhan University, Wuhan 430072, China

[3]Institute for Advanced Studies, Wuhan University, Wuhan 430072, China

*Y.Y. and J.L. contributed equally to this work

†Corresponding author. Email: phxqhuang@scut.edu.cn; dengwy@scut.edu.cn; zyliu@whu.edu.cn

**Abstract**: Topological phases, including the conventional first-order and higher-order topological insulators and semimetals, have emerged as a thriving topic in the fields of condensed-matter physics and material science. Usually, a topological insulator is characterized by a fixed order topological invariant and exhibits associated bulk-boundary correspondence. Here, we realize a new type of topological insulator in a bilayer phononic crystal, which hosts simultaneously the first-order and second-order topologies, referred here as the hybrid-order topological insulator. The one-dimensional gapless helical edge states, and zero-dimensional corner states coexist in the same system. The new hybrid-order topological phase may produce novel applications in topological acoustic devices.

The discovery of topological insulators (TIs), which are characterized by the boundary states in the bulk band gaps, has led to a fundamentally new paradigm in the field of condensed-matter physics [1,2]. The conventional $d$-dimensional ($d$D) TIs, as the first-order TIs, are featured with the $(d-1)$D boundary states [3,4]. For instance, the 2D quantum spin/valley Hall insulators host a pair of 1D gapless helical edge/interface states [5-7]. Recently the higher-order TIs extend the bulk-boundary correspondence, i.e., a $d$D nth-order TIs possess the $(d-n)$D boundary states [8-12]. The 2D second-order TIs with 0D corner states were first predicted [11,12] and have been realized in several systems, including the artificial electronic materials [13], photonic crystals [14-18], phononic crystals (PCs) [19-23] and electric circuits [24-26]. The 3D second-order TIs with 1D hinge states have been achieved in bismuth [27] and acoustic crystals [28].

An interesting question is whether the first-order and higher-order topological properties can occur in a single system, referred as the hybrid-order topological phase. In 2D case, Zhang et al. realized the 1D interface states and 0D corner states in two bulk band gaps, corresponding to the dipole and quadrupole topologies respectively, in a nonsymmorphic metacrystal [29]. In a 3D insulator phase, Kooi et al. proposed that the 2D gapless surface states and 1D hinge states can coexist in a bulk gap by breaking the translational symmetry in a weak TI [30]. Interestingly, the higher-order topological Weyl semimetals [31-33], proposed and realized very recently, also exhibit the hybrid-order features. They not only host the Weyl points and Fermi arcs, as the first-order properties, but also have the 1D hinge states connected the Weyl points originated from the second-order topology [33].

Here we realize a new type of hybrid-order TI in a 2D bilayer PC, which possesses the 1D gapless helical edge states and a pair of 0D corner states localized at one corner in two complete band gaps. Benefiting from no fermionic band-filling for sound waves, the states in different bulk gaps can be measured flexibly in the PC. This hybrid-order TI is composed of a bilayer breathing kagome lattice with chiral interlayer couplings, in which the monopole and dipole modes exhibit the first-order and second-order topologies, respectively. The topology of the monopole mode bands is characterized by the spin-Chern numbers, and leads to the 1D gapless helical edge states in the monopole band gap, as the feature of the first-order TI. But the topology of the dipole mode bands is described by the pseudospin-polarized Wannier centers, with zero spin-Chern numbers, and gives rise to a pair of 0D corner states at one corner in the dipole band gap, as the property of the second-order TI. These results, both the edge states and the corner states, have been verified numerically and experimentally with good agreement.

The lattice model of the bilayer breathing kagome lattice from the top view is plotted in Fig. 1(a), where the green shaded area denotes one unit cell composed of three nonequivalent sites in each layer, labelled as A, B and C. The blue and yellow lines in each layer represent different intralayer couplings, while the dashed arrows denote the interlayer couplings from the lower layer to the upper one. Corresponding to the shaded area in Fig. 1(a), the unit cell of the PC is shown in Fig. 1(b). Each layer of the PC contains three nonequivalent cylinders, as the sites of the lattice model, with the radii and heights being $r = 6.1$ mm and $h = 23.5$ mm, respectively. The lattice constant of each layer is $a = 50$ mm. The diameters of the tubes for the intralayer couplings are $d_1 = 3.2$ mm and $d_2 = 4.96$ mm, and that of the chiral tube for the interlayer coupling is $d_c = 6.6$ mm. The separation between the two layers is $h_1 = 4.7$ mm.

The bulk band dispersion of the PC along the high symmetry lines is calculated in Fig. 1(c). Two complete band gaps, marked in purple and cyan colors, appear in a large frequency separation. To reveal the multipole features of these two gaps, we simulate the eigenmodes at the K points (labelled as $S_1$ and $P_1$), in the bands below these two gaps, as shown in Fig. 1(d). The pressure field distributions of the $S_1$ and $P_1$ modes are derived from the monopole and dipole modes of the cylinder, respectively. Thus, the lower (upper) gap is resulted from the couplings of the monopole (dipole) modes.

We first concentrate on the study of the monopole band gap. For the monopole mode, the cylinders play a role of the lattice sites. They interact with each other through the alternative thick and thin tubes in each layer, and couple with those in the other layer via the chiral tubes. As a result, the real PC for the monopole mode can be described by a tight-binding model, in which the Hamiltonian is written as:

$$H_k = \begin{pmatrix} h_{\text{intra}} & h_{\text{inter}} \\ h_{\text{inter}}^\dagger & h_{\text{intra}} \end{pmatrix}, \tag{1}$$

where $h_{\text{intra}}$ is the Hamiltonian of the breathing kagome lattice in each layer, and $h_{\text{inter}}$ is the interaction of the two layers. By applying a unitary transformation $U = \frac{1}{\sqrt{2}}\begin{pmatrix} 1 & -i \\ 1 & i \end{pmatrix} \otimes I_3$ with $I_3$ the $3 \times 3$ unit matrix, the Hamiltonian $H_u = U H_k U^\dagger$ is deduced as

$$H_u = \begin{pmatrix} h_{\text{intra}} + h_{\text{soc}} & h_{\text{ssc}} \\ h_{\text{ssc}}^\dagger & h_{\text{intra}} - h_{\text{soc}} \end{pmatrix}, \tag{2}$$

where $h_{\text{soc}}$ has the form of the intrinsic spin-orbit coupling, while $h_{\text{ssc}}$ is the coupling between the spin-up and spin-down components. The bulk band dispersion of the tight-binding model [34], described by the parameters $h_{\text{intra}}$, $h_{\text{inter}}$, $h_{\text{soc}}$, and $h_{\text{ssc}}$, is consistent with the monopole mode bands of the PC in Fig. 1(c). Therefore, by introducing a layer pseudospin degree of freedom, the effective intrinsic spin-orbit coupling, as the essential ingredient of the conventional TIs, can be induced by the chiral interlayer tubes, and gives rise to the non-trivial first-order topological properties.

The topology of the monopole mode bands can be described by the spin-Chern numbers [35-38], which have been employed to identify the 2D first-order TIs, including the quantum spin Hall insulators with [37] or without [38] time-reversal symmetry. The spin-Chern numbers of the lowest two monopole mode bands for the PC are calculated as $C_\pm = \pm 1$ [34]. Nonzero spin-Chern numbers guarantee the existence of helical edge states in the monopole band gap [39]. To demonstrate this conventional bulk-boundary correspondence, we fabricate a parallelogram-shaped PC sample, as illustrated in Fig. 2(a), in which the red and green boxes show the enlarged views of the different configurations of boundaries, i.e., the whole-cell and half-cell boundaries, respectively. A point source is positioned on the boundary to excite acoustic waves and a detector is applied to record the pressure field distributions. Through the Fourier transformation of the pressure fields, the projected band dispersions can be obtained. Figures 2(b) and 2(c) show the edge state dispersions along the $k_x$ direction for the whole-cell and half-cell boundaries, respectively. The experimental data are denoted by the color maps, and the full-wave simulated results are represented by the white lines. A pair of counterpropagating gapless edge states exist in the gap for both boundaries. The solid and dashed forms of the white lines denote the opposite spin polarizations along the $y$ direction. So the pair of edge states are helical with spin-momentum locking on each boundary, showing the robustness of the edge states [34].

Subsequently, let us consider the topological property of the dipole band gap. The spin-Chern numbers are also calculated for the two bands below this gap, whereas equal to zero. This indicates that the gap does not host the topological property of the first-order TIs, and is in the absence of the gapless helical edge states. What about the second-order topological property of this gap? By developing the 2D Wannier center to the spin-polarized ones, similar to the procedure of the spin-Chern numbers from the Chern number, we can obtain the second-order topological invariant of the two bands below the dipole band gap, i.e., the spin-polarized 2D Wannier centers, $W_{\pm} = (W_{\pm}^x, W_{\pm}^y) = (-1/2, -1/2\sqrt{3})$ [34]. The nonzero 2D Wannier centers predict the existence of the corner states in a finite sample with its boundary passing through the Wannier center [21]. Because both $W_+$ and $W_-$ are nonzero, there should exist a pair of corner states at a corner of a finite sample.

We then construct the finite sample to demonstrate this second-order bulk-boundary correspondence. The triangle-shaped lattice with 5 unit cells of side length is illustrated in the left panel of Fig. 3(a), and the corresponding PC sample is in the right panel. Note that the cylinders on the boundaries are of whole sizes, same to the triangle-shaped kagome PC of single layer [20,21]. The eigenfrequencies of the triangle-shaped PC are calculated in Fig. 3(b). The red, blue and black circles denote the corner, edge and bulk states, respectively. Six corner states emerge in the gap, as expected. To confirm these eigenmodes, we measure the pressure field distributions from the lower to higher frequencies, which are obtained by exciting and measuring in each cylinder simultaneously. As shown in Fig. 3(c), these field distributions are indeed corresponding to the bulk, edge, corner and bulk states, respectively. It is obviously found that the pressure fields are localized at the three corners at 7.22 kHz. We further observe the acoustic response spectra of the bulk, edge and corner modes. As shown in the top panel of Fig. 3(d), a peak of the response spectrum occurs in the bulk gap, indicating the existence of the corner states. And peaks appear in the regions of the edge and bulk state dispersions because of the existences of the edge and bulk states, as shown by the blue and black lines in the bottom panel of Fig. 3(d).

Finally, we demonstrate a pair of the corner states located at one corner, which is a unique property of the second-order topology in this PC. Because of the $C_2$ symmetry, each corner state should be one of the eigenstates of the symmetry, i.e., state of even or odd parity. As shown in Figs. 4(a) and 4(b), the pair of the corner states possess odd and even parities, respectively. To experimentally distinguish the corner states, two sources with the in-phase or anti-phase are utilized to excite the corner states. We first measure the phase distributions of the cylinders in the upper and lower layers in Figs. 4(c) and 4(d), which show undoubtedly the in-phase and anti-phase features of the corner states. Then we observe the acoustic response spectra with the above two sources. As shown in Fig. 4(e), two peaks of the response spectra occur and coincide around the frequency of the corner states, indicating the coexistence of these two distinct corner states with opposite phase features.

In summary, we have realized an acoustic hybrid-order TI with coexisting of the first-order and second-order topologies for the monopole and dipole modes, respectively. The 1D helical edge states and 0D corner states in two gaps for these modes are observed evidently. In contrast to the electronic systems, in which simultaneous measurements of the edge and corner states in two bulk gaps are difficult due to the fermionic band-filling, our PC provides a powerful platform to explore the hybrid-order topological properties. Moreover, our system may pave a way for innovative applications of acoustics, such as the frequency-dependent functional device, which exhibits the topological waveguide transport for lower frequency and high-performance sensing for higher frequency.

**Acknowledgements**

The authors thank Feng Li for the helpful discussions about the experimental sample.

This work is supported by the National Natural Science Foundation of China (Nos. 11890701, 11804101, 11704128, 11774275, 11974120, 11974005, 12074128), the National Key R&D Program of China (No. 2018YFA0305800), the Guangdong Basic and Applied Basic Research Foundation (No. 2019B151502012), the Guangdong Innovative and Entrepreneurial Research Team Program (No. 2016ZT06C594).

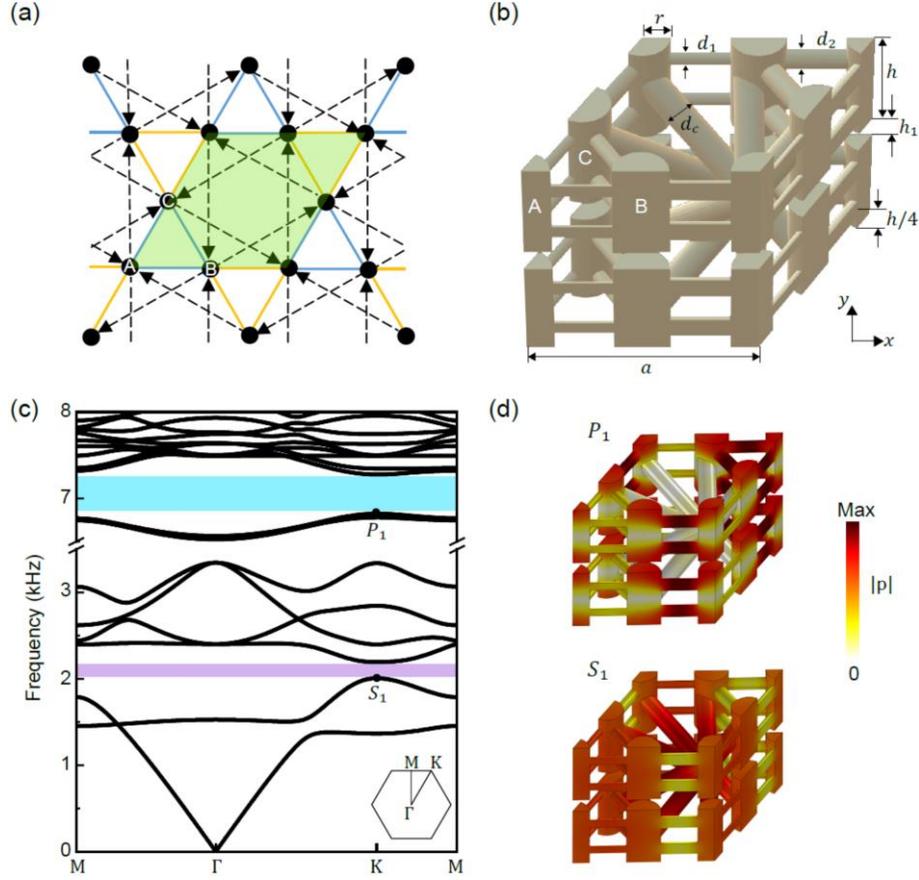

FIG. 1. Acoustic hybrid-order topological insulator with the bulk band dispersions. (a) The model based on the bilayer breathing kagome lattice. The blue and yellow lines denote the different intralayer interactions, and the dashed arrows denote the interlayer couplings from the lower layer to the upper one. (b) The unit cell of the bilayer PC, corresponding to the unit cell marked by the green shaded area in (a). (c) The simulated bulk band dispersions of the PC along the high symmetry lines. Two complete band gaps labelled in purple and cyan colors exhibit the topological properties of the first-order and the second-order TIs, respectively. Inset: the first Brillouin zone. (d) The pressure field distributions of the eigenmodes at the K point below the two band gaps, marked as $S_1$ and $P_1$. They are derived from the monopole and dipole modes of the cylinder, respectively.

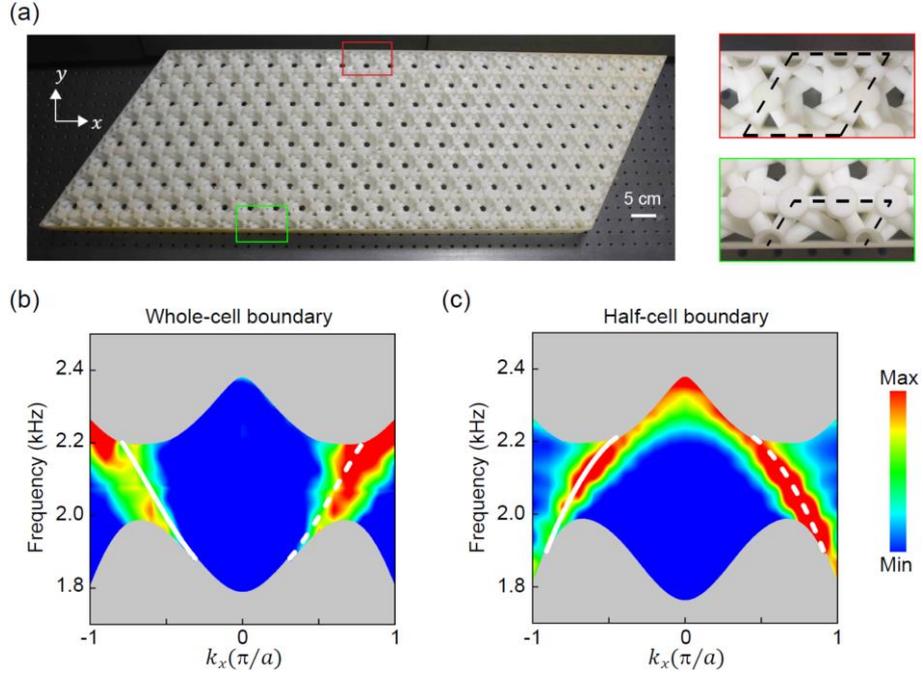

FIG. 2. Acoustic helical edge states in the monopole band gap. (a) Photo of the parallelogram-shaped PC. Red and green boxes denote the enlarged views of the whole-cell and half-cell boundaries (marked by the dashed black lines). (b)-(c) The edge state dispersions on the whole-cell (b) and half-cell (c) boundaries. Color maps represent the measured results, the solid and dashed white lines denote the simulated ones with the opposite spin polarizations. The projected bulk bands are shaded in gray.

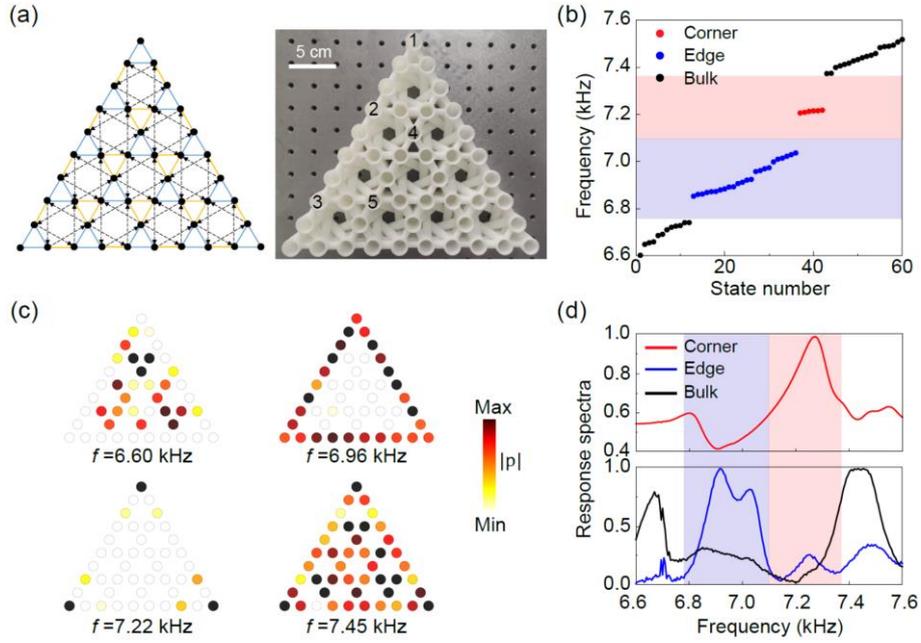

FIG. 3. Acoustic corner states in the dipole band gap. (a) The lattice model (left panel) and PC sample (right panel) of a triangle-shaped structure. (b) The simulated eigenfrequencies of the triangle-shaped PC. Black, blue and red circles represent the bulk, edge and corner states, respectively. (c) The measured pressure field distributions of the bulk (6.60 and 7.45 kHz), edge (6.96 kHz), and corner (7.22 kHz) states. (d) The measured corner (top panel), bulk and edge (bottom panel) response spectra, normalized by their maximum values.

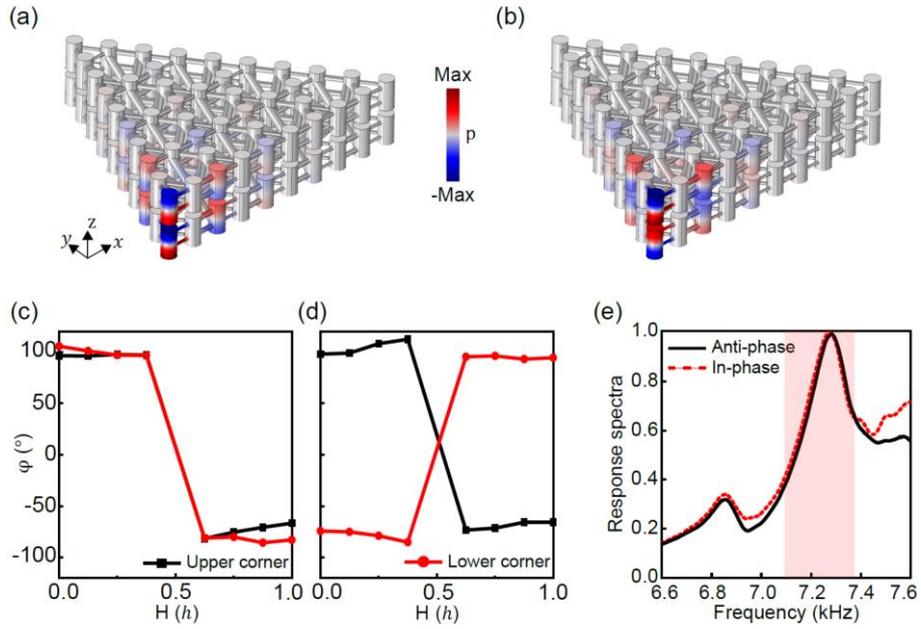

FIG. 4. A pair of acoustic corner states localized at one corner in the dipole band gap. (a)-(b) The simulated eigenmodes at one corner with in-phase and anti-phase distributions, respectively. (c)-(d) The measured acoustic field phases in the cylinders for the upper and lower corners along the $z$ direction, which are consistent with (a) and (b). (e) The measured response spectra of the upper corners with respect to the in-phase and anti-phase distributions.